\documentclass{iopart}

\input epsf
\usepackage{graphicx}
\usepackage{cite}

\usepackage{bm}
\usepackage{mathrsfs}
\usepackage{amssymb}
\usepackage{stmaryrd}

\newcommand{\eq}[1]{Eq.~(\ref{#1})}

\newcommand{\be}[1]{\begin{equation}\label{#1}}
\newcommand{\ee}{\end{equation}}
\renewcommand{\vec}{\mathbf}

\begin{document}

\jl 2
\letter{Triple photoionization of Lithium near threshold}
\author{Agapi Emmanouilidou\dag\ and Jan M.\ Rost\ddag}
\address{\dag Center for Nonlinear Science, School of Physics, Georgia Institute of Technology, Atlanta, Georgia, 30332-0430}
\address{\ddag Max Planck Institute for the Physics of Complex Systems, D-01187 Dresden, Germany}
\date{\today}
\begin{abstract}
Solving the full classical four-body Coulomb problem numerically
using a Wigner initial distribution we formulate a classical-quantum hybrid approach to study triple ionization by single photon absorption
from the Li ground state in the threshold region. 
 We confirm the Wannier threshold law $\sigma \propto E^{\alpha}$ and we
show that the $\alpha$ determined in the interval between $2-5$
eV deviates from the analytical threshold value of 2.16 which we find in
the interval between $0.1-2$ eV.  
\pacs{3.65.Sq, 32.80.Fb, 34.80.Dp}
\end{abstract}

\noindent
Triple photoionization of Lithium is the most fundamental atomic process
involving three bound electrons. Only recently Wehlitz {\it et al.}
succeeded in measuring the triple photoionization cross section
down to 2 eV above threshold \cite{Wehal98,Wehal00}. Subsequent experiments have produced various double ionization cross sections,
also very close to threshold.  For Lithium \cite{Wehal02} and most recently
Beryllium \cite{Wehal04}, it was demonstrated convincingly that
the double photoionization cross section has  small oscillations superimposed on the rising smooth cross section. This has cast some doubt on the validity of Wannier's (classically derived) threshold law \cite{Wan53,Wan55} which predicts very close to threshold a power law
behavior of the cross section, 
\be{wannier}
\sigma(E_{\omega}) \propto \left(E_{\omega}/I-1\right)^{\alpha},
\ee
where $E_{\omega}$ is the photon energy, $I$ the respective threshold energy
and $\alpha$ a characteristic exponent which is related to the stability of a classical fixed point of the $N$-electron dynamics \cite{Ros01}. 

For the present case of Lithium triple ionization ($I=7.478$ a.u.) the experimental results
reach only down to 2 eV above threshold \cite{Wehal00}. The corresponding fit
of the experimental data with \eq{wannier} yields $\alpha_{\mathrm{exp}}=2.05$ close to the theoretical value of $\alpha = 2.16$
derived by Klar and Schlecht \cite{Klar}. It is known from the well studied
two-electron escape case \cite{Ros94} that a Wannier exponent fitted to a
cross section over an energy interval which is close but a finite distance
away from threshold yields always a {\it smaller} $\alpha$ than the
analytically predicted one. Hence, the experimental value is consistent with Wannier's prediction
but not conclusive since one does not know what happens closer to threshold.

The present classical study deals for the first time with the full four-body problem close to threshold and provides the triple photoionization (TPI) probability 
starting at $E=0.9$~eV excess energy. We are able to confirm that, indeed, the Wannier threshold law with an exponent of 
$\alpha = 2.16$ is reached, but only for energies $E\equiv E_{\omega}-I< 2$eV. By successively fitting finite energy intervals above the threshold $I$ to our result, we can confirm the experimental result for $\alpha$.

We formulate the TPI process from the Li ground state ($1s^{2}2s$) as a two step process \cite{Samson, Pattard}. First, one electron absorbs the photon (photo-electron). Then, due to the electronic correlations, redistribution of the energy takes place resulting in three electrons escaping to the continuum. We express the above two step process as
  \begin{equation}
\label{prob}
\sigma^{3+} = \sigma_{\rm abs} P^{3+},\
\end{equation}  
 where $\sigma_{abs}$ is the total absorption cross section and $P^{3+}$ is the probability for triple ionization. In what follows, we evaluate $P^{3+}$ and use the experimental data of Wehlitz \cite{total} for $\sigma_{abs}$. 
 Physically interpreted, this relation splits the photon absorption ($
 \sigma_{\rm abs}$) from the subsequent energy redistribution in the
 three-electron system. The latter can lead to TPI which we calculate
 in phase space formally from
\begin{equation}
P^{3+}=\lim_{t\rightarrow\infty}\int{\rm d}\Gamma_{\mathcal
{P}^{3+}}\,
\exp((t-t_{{\rm abs}})\mathscr{L}_{\mathrm{cl}})\rho(\Gamma),
\label{intphas}
\end{equation}
with the classical Liouvillian
$\mathscr{L}_{\mathrm{cl}}$ given by the Poisson bracket \{H, \}
\cite{Henriksen} propagated from the time
$t_{\mathrm{abs}}$ of photo absorption, with H the four-body Coulomb Hamiltonian. 
 The primary electron absorbs the photon at the nucleus
($\vec r_{1}=0$),
 an approximation which becomes exact in the limit of high photon
energy \cite{kabir}. We note that no account is taken of the direction of polarization of the incident photon, since we currently consider electron orbitals 
that are spherically symmetric. Immediately after absorption, the phase space distribution of the remaining two electrons 
is the Wigner transform of the corresponding initial wavefunction
$\psi(\vec r_{1}=0,\vec r_{2},\vec r_{3})$, where the $\vec r_{i}$ are the
electron vectors starting at the nucleus. 

 In general, close to threshold the ionization probability
does not depend on the details of the initial wavefunction \cite{Wan53}. Hence, we approximate it as a simple product of hydrogenic orbitals $\phi^{\mathrm{Z}_{i}}_{i}(\vec r_{i})$ with effective charges $Z_{i}$  to facilitate the Wigner transformation. The $Z_{i}$ are chosen to reproduce the known ionization potentials $I_{i}$,
namely for the 2s electron $Z_{3}=1.259$ ($I_{3}=0.198\,$a.u.) and for the 1s electron $Z_{2}=2.358$ ($I_{2}=2.780\,$a.u.). We use atomic units throughout the paper if not  stated otherwise. The Wigner distribution $W$ conserves energy only in the mean \cite{GeRo02}. Near $E = 0$, however,  energy conservation is vital. Therefore, the Wigner distributions for the individual electron orbitals $\phi^{\mathrm{Z}_{i}}_{i}(\vec r_{i})$, $W_{\phi^{\mathrm{Z}_{i}}_{i}}$, are restricted to their respective energy shell, leading to the initial phase space distribution
\be{wigtrans}
\rho(\Gamma) =  \mathscr{N} \delta(\vec{r}_1)\delta(\varepsilon_{1}+I_{1}-\omega)\prod_{i=2,3}W_{\phi^{\mathrm{Z}_{i}}_{i}}(\vec r_{i},\vec p_{i})\delta(\varepsilon_{i}+I_{i})\,,
\ee
where $\epsilon_{i}$, $i=1,2,3$ are the individual electron energies. The advantages of the Wigner distribution as an initial phase space distribution over other distributions as 
well as the reasons for restricting the Wigner distribution of the hydrogenic orbitals on their respective energy shell is discussed in ref. \cite{CTMC3}.

\begin{figure}
\centerline{\includegraphics[scale=0.6,clip=true]{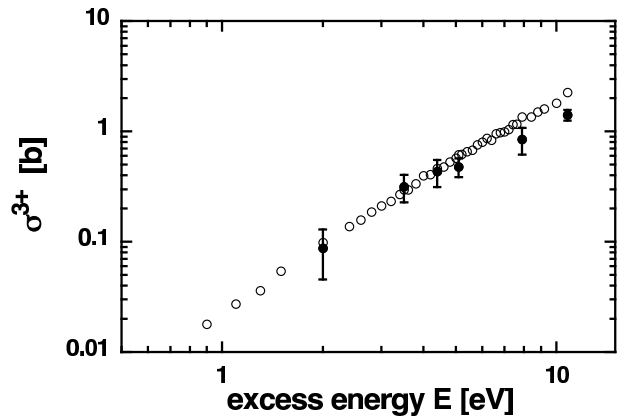}}
\caption{\label{fig1:li-threshold} TPI cross section
obtained by multiplying the TPI probability  from the present
calculation with the total photo cross section from \cite{total} 
($\circ$) in comparison to the experiment \cite{Wehal00} ($\bullet$).}
\end{figure}

With $\rho(\Gamma)$ from \eq{wigtrans}  the
initial phase space volume to be sampled reduces significantly, although 
regularized coordinates
\cite{ks1} are required to avoid problems with electron trajectories
starting at the nucleus. Other than that, the integral in Eq.~(\ref{intphas}) is evaluated with a
standard Monte-Carlo technique which entails following classical trajectories in phase space (CTMC) \cite{CTMC1,CTMC2,CTMC3,CTMC4}. The projector $\mathcal P^{3+}$ indicates that we integrate only over those parts of phase space that lead to TPI. Triple ionization is decided by propagating trajectories long enough so
that the individual electron energies $\epsilon_{i}$, $i=1,2,3$, are
positive and stable within a margin that guarantees all three electrons
are very far away from the nucleus and each other. A similar
approach to the one described above was successfully used to describe the
knockout mechanism for the double
ionization of He from the ground \cite{Schal02}, the $2^{1,3}S$ excited
states \cite{Agapi}, and the double ionization of H$_{2}$ \cite{Christian}.

 We also note that generally, not only close to threshold, the classical TPI trajectories provide information as to how energy is redistributed from the primary photo electron to the other two electrons 
 which is analyzed in detail in \cite{AgapiPRL}.

\begin{figure}
\centerline{\includegraphics[scale=0.9,clip=true]{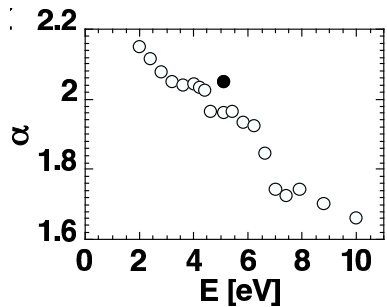}}
\caption{\label{fig1:exponent}
 The Wannier exponent $\alpha$, obtained by fitting  finite energy intervals to \eq{wannier}, see text. The filled circle is the experimental value from \cite{Wehal00}.}
\end{figure}

Figure \ref{fig1:li-threshold} shows the TPI cross section $\sigma^{3+}$ resulting from the probability $P^{3+}$
in connection with \eq{prob}. We find very good agreement with the
experimental results. Thus, our classical approach with an approximate initial
quantum wavefunction captures the relevant correlations among the three electrons which
mainly form after the photo absorption at lower excess energies. Let us note that the reason we currently consider energies of 0.9 eV and above is the numerical difficulty involved in computing $P^{3+}$. Specifically, in order to obtain $10^{3}$ TPI trajectories at $E=0.9$ eV one has to evolve $10^{10}$ trajectories with the CTMC method.  
 One may be tempted to see a slightly different slope of the experimental
curve compared to the theoretical one in Figure
\ref{fig1:li-threshold}. Concerning the theoretical curve this may be due
to the fact that we employ a ``high energy'' approximation for the photo
absorption,
 namely that the photon is absorbed by an electron that is sitting initially
at the nucleus ($\vec r_{1}=0$). In the future we may relax this approximation 
by using a photon-frequency dependent assumption for $r(\omega)$ of the photon-electron
as described in ref.\cite{Rost10}. However, given the present accuracy of the
experiment near threshold the slope of the experimental curve is somewhat
uncertain as well, see error bars in Figure \ref{fig1:li-threshold}.
 Luckily these experimental and theoretical difficulties do not hamper the
present goal of analyzing the behavior of the cross section towards threshold,
$E\rightarrow 0$. We want to investigate whether the
 Wannier power law for $\sigma^{3+}$ is really approached in the limit of vanishing threshold energy.
To this end we have fitted $\sigma^{3+}= \sigma_{0}(E/I)^{\alpha}$, where $\sigma_{0}$ and $\alpha$ are fit parameters while $I$ is the triple ionization potential. For the closest energy interval to threshold we could reach, 0.9 eV$\,\le E \le\,$2 eV, $\alpha_{\mathrm{theo}} = 2.15$ very close to the analytical value of $\alpha = 2.16$, 
as Figure \ref{fig1:exponent} reveals. We then apply the fit, keeping the lower limit of
the energy interval constant, $E=0.9$ eV, 
and increasing the upper energy limit until we reach $E=4.0$ eV ($\alpha_{\mathrm{exp}}$ is obtained in the range $2-5.1$ eV $=3.1$ eV). Subsequently,
 we shift the $3.1$ eV interval to higher excess energies to
obtain $\alpha$ as a function of the upper limit of the energy range
\cite{Wehal04a}.

 Figure \ref{fig1:exponent} illustrates that the Wannier exponent fitted to the
cross section over an energy interval which is close but a finite distance
away from threshold yields a {\it smaller} $\alpha$ than the
analytically predicted one. Hence, the fit to the experimental
data (filled circle in Figure \ref{fig1:li-threshold}) is consistent with Wannier's theory and in good agreement with our present theoretical result. The strong variation of the Wannier exponent in  figure \ref{fig1:exponent} also indicates that the threshold region where Wannier's threshold law applies is certainly less than the energy range shown in 
 Figure \ref{fig1:exponent}.

Secondly, we explore if the triple photoionization cross section for different
atomic elements have a similar shape close to the threshold region. We use the shape formula for the TPI cross section \cite{Pat02},
\begin{equation}
\label{eq:shape}
\sigma^{3+}=\sigma_{M}x^{\alpha}\left(\frac{\alpha+7/2}{\alpha
x+7/2}\right)^{\alpha+7/2},
\ee
to obtain  a dimensionless cross section  $\sigma/\sigma_{M}$ as a function of
the dimensionless excess energy $x=E/E_{M}$ with $E_{M},\sigma_{M}$ as fitting
parameters and with $\alpha$ set to its analytical value of 2.16. 
 Eq.(\ref{eq:shape}) reproduces, by construction, for Coulomb complete
break-up processes, the Wannier threshold law
and the cross section for high excess energies. In Figure
\ref{fig6:triple-scaledcs}, one sees that the experimental data for  Li, Ar, and Ne
fall on top of the theoretical data from Figure \ref{fig1:li-threshold}.
\begin{figure}
\centerline{\includegraphics[scale=0.6,clip=true]{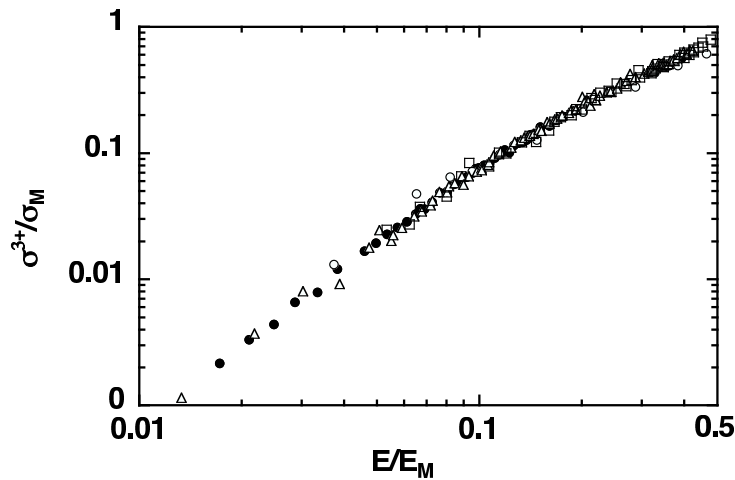}}
\caption{\label{fig6:triple-scaledcs}  TPI cross section in scaled coordinates
\cite{Pat02} for Lithium  (this work, $\bullet$), and from the experiments
 on Lithium (\cite{Wehal00}, $\circ$), Neon (\cite{Wehal04a}, $\bigtriangleup$)
and Argon (\cite{Wehal04a}, $\square$).}
\end{figure}

It has been argued that a secondary power law, or at least additional structure in the TPI cross section could originate from the very different binding energies for the electrons in the Lithium atom with its $1s^{2}2s$ configuration in contrast to Neon and  Argon which both contribute three electrons from a single shell to the ionization (2p and 3p, respectively)  \cite{Wehal00}.
 The present calculation does take into account the difference in
binding energy and spatial extension
of the respective orbitals. We find our results to be consistent with a smooth change of the TPI cross section
for the ground state of Li.
 Hence, one may conclude from the agreement of our Lithium calculation regarding the shape of the ionization cross section with Neon and Argon that the different binding energies do not strongly influence the shape of the cross section.

In summary in the framework of a quantum-classical hybrid approach we have analyzed triple photoionization of Lithium near
threshold using a Wigner initial distribution and a classical propagation of the three electrons in time. We can confirm 
that the total cross section grows from threshold with a power of $\alpha =
2.16$ in accordance with Wannier's threshold theory. In addition, using
the shape formula, we find our results to be consistent with a smooth
change of the triple ionization cross section for the Li ground state.

We gratefully acknowledge discussions with Thomas Pattard and thank Ralf Wehlitz for providing his data in electronic form.\\

{\bf References}\\

\end{document}